# *Low-loss phase-change material based programmable mode converter for photonic computing*

Xueyang Shen[1], Ruixuan Chu[1], Ding Xu[1], Yuan Gao[1], Wen Zhou[1*], Wei Zhang[1*]

[1]Center for Alloy Innovation and Design (CAID), State Key Laboratory for Mechanical Behavior of Materials, Xi'an Jiaotong University, Xi'an, 710049, China.

*Email: wen.zhou@xjtu.edu.cn, wzhang0@mail.xjtu.edu.cn

**Abstract**

Phase-change materials (PCMs)-based integrated photonic memory offers a viable pathway for the development of neuromorphic computing chip. The sizable optical contrast in the telecom band between amorphous and crystalline phases of PCM, in particular, $Ge_2Sb_2Te_5$ (GST), is used for multilevel programming. However, the high extinction coefficient *k* of crystalline GST leads to high optical loss, posing a serious challenge for scaling up the device array for practical use. In this work, we focus on the atomic understanding and application of the so-called low-loss PCM, $Sb_2Se_3$, through multiscale simulations. First, we elucidate the bonding origin of the wavelength dependent optical properties of amorphous and crystalline $Sb_2Se_3$ via *ab initio* calculations. Given the suppressed *k* in the telecom band, we design a programable mode converter (PMC) waveguide device that utilizes only the contrast in refractive index *n* between amorphous and crystalline $Sb_2Se_3$ to encode multiple optical levels per waveguide device. The finite-difference time-domain simulations show that a single PMC device can achieve 5-bit programming precision (32 levels) via direct laser writing, and the photonic tensor core formed by the PMC array could possibly be scaled to 128×128. Finally, a thorough comparison between low-loss PCM and conventional PCM is provided.

**Keywords:** phase-change materials, optical properties, low-loss, multiscale simulation, photonic computing

**Introduction**

The global demand for data storage and processing is growing drastically, yet the associated power consumption has become unsustainable. There is an urgent need to develop low-power, non-volatile, and small-footprint memory and neuromorphic computing devices[1-6]. Phase-change materials (PCMs) utilize the reversible phase transition between the amorphous (*a*-) and crystalline (*c*-) phases and the associated changes in electrical and optical properties to storage data[7-13]. The flagship Ge-Sb-Te alloys have been commercialized for both high-density stand-alone and high-stability embedded memory products[14-17]. By fine-tuning the amorphous-to-crystal ratio, it is feasible to generate multiple resistance levels in a single memory cell, supporting efficient implementation of neuromorphic computing functions using memory arrays[4]. Using doped $Ge_2Sb_2Te_5$ (GST) phase-change alloy, prototype neuromorphic computing chip was shown to be capable of resolving many demanding artificial intelligence (AI) tasks at high computing and energy efficiencies[18-20]. The prototype chip consists of multiple computing cores, and each core contains a 256×256 memory array[4]. Owing to the resistance drift issue, the computing precision is still limited, and the new materials with low-drift characters are still being pursued[21].

By integrating GST thin films with silicon-on-insulator (SOI) waveguides, it is feasible to fabricate photonic array of high density[22-24]. PCM-based photonic neuromorphic computing is under active development[25-30]. The all-optical photonic computing scheme exploits light-matter interactions between GST thin films and lasers in the telecom bands, e.g. 1550 nm, to trigger phase transition. The sizable contrast in light transmittance is used for memory programming. In comparison with its electronic analogue, PCM photonic computing devices offer much larger bandwidth for data transfer and the parallel computing capability using the wavelength-division-multiplexing (WDM) scheme[31,32]. Despite successful demonstrations in achieving the convolutional operation[33,34], matrix-vector multiplication[35,36], and tensor processing functions[37,38], the core size of the photonic GST array is limited to 3×3 for practical use, which is mainly due to the high optical loss of crystalline GST. Feldmann *et al.* demonstrated experimentally that weak light signals could still be detected after passing through a crossbar array with 32 GST cells per row[33]. In parallel, Wu *et al.* patterned the GST thin film into a metasurface, and developed a multimode programming scheme for photonic computing, which could in principle support 45 cells per row[34].

The high optical loss (strong electronic excitation) of crystalline GST stems from two factors, namely, a narrow band gap of well below 1 eV and a special bonding mechanism, termed as metavalent bonding (MVB)[39-43]. Upon rapid crystallization, GST forms a rocksalt-like structure with well-aligned *p* orbitals, which leads to high probability of inter-band transitions, as evidenced by the high values of transition dipole moment[44]. This strong optical loss limits the size and scalability of the GST-based photonic device array. A material solution for overcoming the performance limitation is to substitute Te with Se elements[45], e.g., $Ge_2Sb_2Se_4Te_1$ (GSST) to reduce the optical loss in the telecom band[45-48]. Moreover, the binary alloy $Sb_2Se_3$[49-51] shows nearly no light absorption at 1550 nm for both phases, but a non-negligible change in refractive index *n* of 0.77 is preserved upon phase transition[52], making $Sb_2Se_3$ a promising candidate for optical applications, including tunable microring resonator (MRR)[53], Mach-Zehnder interferometers (MZI)[54-56], and multimode interference (MMI) device[57,58].

However, the atomic origin of the low-loss mechanism of $Sb_2Se_3$ is not yet fully clarified, in particular, it remains to be answered why an alloy with covalent bonding character[40] can also be utilized for photonic phase-change applications. Given the limited Δ*n*, how to develop a photonic crossbar array with many optical levels per cell for high programming precision and low optical loss per cell for much extended matrix size simultaneously. Besides, it remains to be explored whether the low-loss PCM based photonic tensor core can support high-performance neuromorphic computing functions. To address these questions, we perform multiscale simulations[59] in the following sections.

**Results**
**Bonding and optical properties of the low-loss PCM $Sb_2Se_3$**
We performed density functional theory (DFT) calculations and DFT-based *ab initio* molecular dynamics (AIMD) simulations to investigate crystalline and amorphous $Sb_2Se_3$. The obtained atomic structures are shown in Fig. 1a. Three independent amorphous models (*a*-), each containing 180 atoms were generated by following the standard melt-quench protocol[60] (see details in the Method section). In the amorphous models, Sb atoms are mostly three-fold coordinated and the major bond angle around Sb atoms is ~90°, suggesting defective octahedral configurations. Further structural analyses of *a*-$Sb_2Se_3$ can be found in Fig. S1, which are consistent with literature data[61]. Upon crystallization, $Sb_2Se_3$ forms an orthorhombic (*o*-) structure[62] with zigzag-shape van der Waals (vdW)-like gap. The nearest interatomic distance across the structural gap ranges from 3.42 Å to 4.00 Å. Within each atomic slab, Sb atoms are six-fold and five-fold coordinated with Se atoms, and Peierls distortion can be found in the Se-Sb-Se bond pairs. The *o*-phase model is approximately 91 meV/atom lower in energy as compared to the *a*-phase models. We also considered two hypothetical phases of $Sb_2Se_3$, taking the crystalline structures of its homologue compounds: the rhombohedral (*r*-) phase of $Sb_2Te_3$, denoted as *r*-$Sb_2Se_3$ (Fig. 1a middle model), and the monoclinic (*m*-) phase of $As_2Se_3$, denoted as *m*-$Sb_2Se_3$ (Fig. S2). The cell volume was relaxed to reduce the internal stresses. For *r*-$Sb_2Se_3$, the vdW-like gaps are ordered with an interatomic distance of 3.58 Å, and each atomic slab consists of five atomic layers with alternating short and long bonds. Note that this *r*-phase model is actually energetically favorable over *o*-$Sb_2Se_3$ by 25 meV/atom, consistent with previous findings[63]. In our previous work, we showed that the degree of MVB can be tailored by applying uniaxial strain perpendicular to the vdW-like gaps for layered PCMs[64]. Here, we also calculated several strained rhombohedral (*s-r*-) models with the c-axis being varied from 0 Å to –2.5 Å. If not specified, the *s-r*-$Sb_2Se_3$ model refers to the mostly strained rhombohedral model. As regards *m*-$Sb_2Se_3$, all the Sb atoms form bonds with three neighboring Se atoms, where Peierls distortion is completely absent. The structural details and energy values of amorphous and various crystalline Sb₂Se₃ models are summarized in Table 1. All the structural files can be found in an open repository (see Data Availability section).

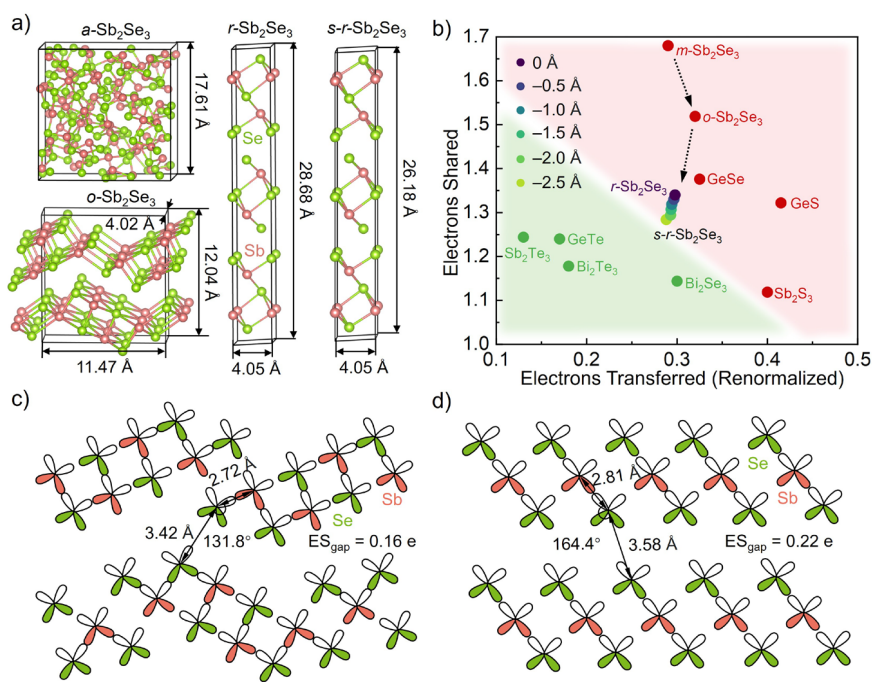

**Fig. 1 Structures and bonding properties of Sb$_2$Se$_3$.** a) The relaxed models of amorphous (*a*-), orthorhombic (*o*-), rhombohedral (*r*-) and strained rhombohedral (*s-r*-) phases of Sb$_2$Se$_3$. The *a*-, *o*- and *r*-Sb$_2$Se$_3$ models contain 180, 20 and 15 atoms, respectively. The Sb and Se atoms are rendered with red and green spheres, respectively. b) The bonding map that includes various crystalline models of Sb$_2$Se$_3$. c-d) Schematic of *p* orbitals arrangement in c) *o*- and d) *r*-Sb$_2$Se$_3$.

| Sb$_2$Se$_3$ | a (Å) | b (Å) | c (Å) | α(°) | β(°) | γ(°) | $E_{diff}$ (meV/atom) |
|---|---|---|---|---|---|---|---|
| *a*- | 17.61 | 17.61 | 17.61 | 90 | 90 | 90 | 115.98 |
| *o*- | 12.04 | 4.02 | 11.47 | 90 | 90 | 90 | 24.90 |
| *m*- | 13.22 | 10.18 | 4.15 | 90 | 90.78 | 90 | 45.83 |
| *r*- | 4.05 | 4.05 | 28.68 | 90 | 90 | 120 | 0 |
| strained *r*- (–0.5 Å) | 4.05 | 4.05 | 28.18 | 90 | 90 | 120 | 1.95 |
| strained *r*- (–1.0 Å) | 4.05 | 4.05 | 27.68 | 90 | 90 | 120 | 4.84 |
| strained *r*- (–1.5 Å) | 4.05 | 4.05 | 27.18 | 90 | 90 | 120 | 10.49 |
| strained *r*- (–2.0 Å) | 4.05 | 4.05 | 26.68 | 90 | 90 | 120 | 19.82 |
| *s-r*- (–2.5 Å) | 4.05 | 4.05 | 26.18 | 90 | 90 | 120 | 33.69 |

**Table 1.** The relaxed lattice parameters and the energy differences of the *a*-, *o*-, *m*-, *r*-, four strained *r*-, and mostly strained *s-r*-Sb$_2$Se$_3$ models. The energy difference ($E_{diff}$) values were calculated with respect to the *r*-Sb$_2$Se$_3$ model.

To gain further understanding of the bonding characters, we calculated the normalized electrons transferred (ET) and electrons shared (ES) values of the crystalline $Sb_2Se_3$ models. The ET values are quantified by normalizing the electrons transferred to the oxidation state, and ES values are twice of the shared electron pairs between adjacent atoms[39] (see Method for details). As shown in the bonding map in Fig. 1b, the green and red area represent the regions of MVB and covalent bonding, respectively. Only the ET and ES values for the short Sb–Se bonds are plotted on the map (Fig. 1b)[65]. The ES values gradually decreased from the *m*-phase to the *o*-phase, and then to the *r*-phase. Uniaxial strain gradually reduced the interatomic distance between the adjacent Se atoms across the structural gaps from 3.58 Å to 3.16 Å in *r*-phase models, which resulted in reductions in both ET and ES values. The strained models gradually moved to the MVB region, and the interaction between the Se atoms was also enhanced. The averaged electrons shared value between two weakly coupled Se and Se atoms across the structural gap, denoted as $ES_{gap}$, was increased from 0.22 e to 0.44 e. Regarding the *o*-phase, the average $ES_{gap}$ value was found as 0.16 e. We also depicted the bonding environment of *o*-$Sb_2Se_3$ and *r*-$Sb_2Se_3$ in term of *p* orbitals (Fig. 1c and 1d). In *o*-$Sb_2Se_3$, four Sb–Se bonds within the atom layers are coupled through the *p* orbital alignment, with bond angles around the Sb and Se atoms ranging from 165° to 175°. The orbital arrangement breaks down across the zigzag vdW-like gaps, e.g., the bond angle is ~131.8° for the Sb–Se⋯Se bond pair that contains the shortest interatomic Se⋯Se contact across the structural gaps. However, in *r*-$Sb_2Se_3$, the *p* orbitals are well-aligned inside all the atomic slabs as well as across the vdW-like gaps. The bond angle in the Sb–Se⋯Se bond pair across the structural gaps is 164.4°, and is improved to 170.7° in the mostly strained model (−2.5 Å). The bonding character, in particular, the degree of *p* orbital alignment directly affects the optical properties.

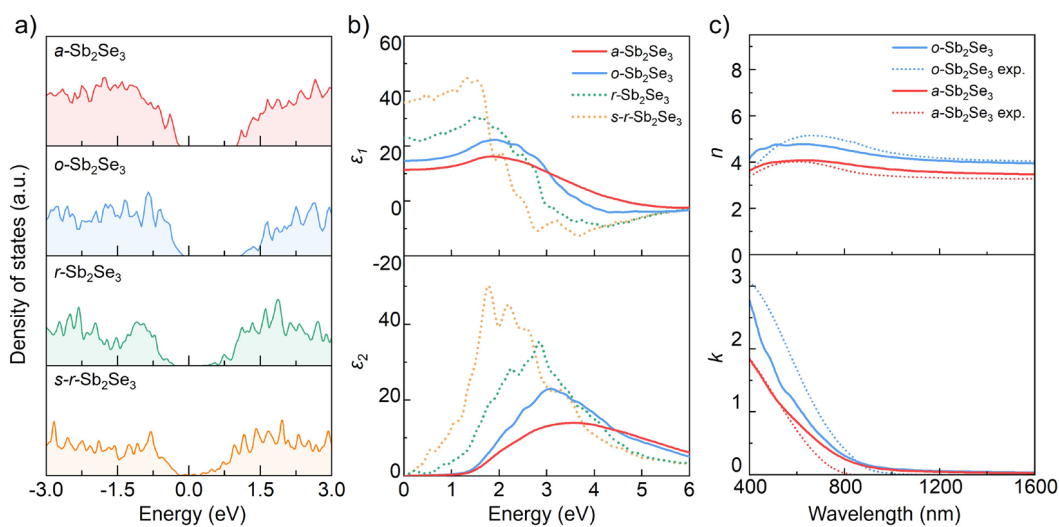

**Fig 2. Electronic and optical properties of $Sb_2Se_3$ calculated by using HSE06 hybrid functional.** a) Density of states of *a*-, *o*-, *r*- and *s*-*r*-$Sb_2Se_3$. b) Real ($\varepsilon_1$) and imaginary ($\varepsilon_2$) parts of the dielectric functions of $Sb_2Se_3$. The results of hypothetical rhombohedral model and corresponding strained model are depicted in dashed lines. c) Calculated and experimental refractive index (*n*) and extinction coefficient (*k*) of $Sb_2Se_3$. The experimental data are adapted from Ref.[52] with permission, Wiley-VCH.

Fig. 2a shows the density of states (DOS) for various $Sb_2Se_3$ models. The band gap values of both a-$Sb_2Se_3$ and o-$Sb_2Se_3$ were calculated as ~1.3 eV, and those of rhombohedral models were smaller. To evaluate the optical properties of $Sb_2Se_3$ models, we performed frequency-dependent dielectric function calculations using the independent-particle approximation[66-68]. Fig. 2b shows the real ($\varepsilon_1$) and imaginary parts ($\varepsilon_2$) of the dielectric functions for a-, o-, r- and s-r-$Sb_2Se_3$. The optical properties of other strained models are shown in Fig. S3. Consistent with the predictions from the bonding map, the s-r-$Sb_2Se_3$ model, which is closest to the MVB region, shows the highest $\varepsilon_1$ value at 0 eV. For all the crystalline models, a significant reduction in the $\varepsilon_2$ profile can be observed below 4.5 eV, thereby a decrease in optical contrast with respect to a-$Sb_2Se_3$. The value of $\varepsilon_2$ is contributed jointly by the amount of possible inter-band excitations (joint density of states, JDOS), and the possibility of transition (transition dipole moments, TDM)[44,67]. As shown in Fig. S4a, the overall JDOS profile for o-$Sb_2Se_3$ and a-$Sb_2Se_3$ are comparable. In the low-energy region below ~1 eV (i.e., the region with wavelengths larger than ~1240 nm), there are no JDOS values for both phases, which is also reflected by the band gap edge above Fermi level in DOS profiles. But the TDM values for o-$Sb_2Se_3$ are consistently higher than those of a-$Sb_2Se_3$, contributing to the optical contrast between two phases. Note that no TDM values can be found below ~1.3 eV, which means that no interband transitions would occur. For r-$Sb_2Se_3$ and s-r-$Sb_2Se_3$ with well-aligned p orbitals, the TDM values are even higher (Fig. S4b). An increased degree of p orbital alignment results in a larger electron delocalization. This leads to a greater wavefunction overlap between the initial states in the valence band and the final states in the conduction band, yielding an enhancement of the optical transition matrix element and thereby optical response. Considering the smaller band gap size of the r-phase models, much higher values are found in the $\varepsilon_2$ profile below 3 eV, resulting in a much larger optical response in both the visible light and near infrared range.

The refractive index (n) and the extinction coefficient (k) can be calculated from the real ($\varepsilon_1$) and imaginary ($\varepsilon_2$) parts of the dielectric functions[69]:

$$n(\omega) = \left(\frac{\sqrt{\varepsilon_1^2 + \varepsilon_2^2} + \varepsilon_1}{2}\right)^{\frac{1}{2}}, \qquad (1)$$

$$k(\omega) = \left(\frac{\sqrt{\varepsilon_1^2 + \varepsilon_2^2} - \varepsilon_1}{2}\right)^{\frac{1}{2}}, \qquad (2)$$

where $\omega$ is the photon frequency. Fig. 2c shows n and k of o-$Sb_2Se_3$ and a-$Sb_2Se_3$. The experimental data from Ref.[52] are also plotted using dashed lines. The calculated optical profiles compare well with the experimental data. At the wavelength range from the visible light (~ 400–800 nm) to the telecom bands (~1500–1600 nm), a clear contrast window of the refractive index ($\Delta n$) between o-$Sb_2Se_3$ and a-$Sb_2Se_3$ can be observed. For the r-phase models, the $\Delta n$ is much larger, in particular, in the strained models (Fig. S4c). The k profiles of both o-$Sb_2Se_3$ and a-$Sb_2Se_3$ decrease to nearly zero at the wavelength of 1550 nm. Since k is determined mostly by $\varepsilon_2$ (Eq. (2)), the low-loss property of $Sb_2Se_3$ originates from the vanished JDOS and TDM values in this wavelength range. A detailed comparison of $k$, JDOS and TDM for the $r$-, $o$- and $m$-$Sb_2Se_3$ models can be found in Fig. S5. It is worth noting that below the wavelength of ~800 nm, a certain degree of absorption is still observed. Therefore, using visible light lasers, it is still feasible to induce sufficient Joule heating via light absorption for phase transition. It is important to employ the Heyd-Scuseria-Ernzerhof (HSE06) hybrid functional[70] for the optical calculations of $Sb_2Se_3$, as such functional leads to a more accurate description of band gap size. As shown in Fig. 2c and Fig. S6, the DFT-HSE06 calculations reproduced experimental observation[52] of negligible optical loss in the telecom band range but finite optical loss in the visible-light range for both amorphous and orthorhombic $Sb_2Se_3$.

Overall, the optical contrast between crystalline and amorphous $Sb_2Se_3$ is wavelength dependent, and is strongly affected by the degree of $p$ orbital alignment. Although not fully ordered, the $p$ orbitals in the $o$-$Sb_2Se_3$ atomic slabs are still partly aligned. If the orbital alignment is completely disrupted, e.g., in $m$-$Sb_2Se_3$, where the local units are in a distorted and defective tetrahedral configuration, the optical response becomes amorphous-like (dotted purple lines versus solid red lines in Fig. S4c). Moreover, the mass density of the $o$-phase is larger than that of the $a$-phase, and the density variation can affect the size of band gap for optical excitations. If $r$-$Sb_2Se_3$ with high degree of $p$ orbital alignment can be produced experimentally, much higher $n$ and $k$ values can be expected. Nevertheless, in such case, the material is no longer expected to display low-loss features in the telecom band.

**Design of the programmable mode converter (PMC)**
Based on the DFT-calculated bonding and optical properties, we elucidate the wavelength-dependent low-loss and finite-loss feature of $Sb_2Se_3$. Both amorphous phase and crystalline $o$-phase of $Sb_2Se_3$ are fully transmissive at the telecom band, but these two phases still show relatively high optical absorptions in the visible-light range, which can be switched reversible via photothermal effects using lasers with a wavelength between 400–700 nm. Since there is no contrast in $k$ between $a$- and $o$-$Sb_2Se_3$ in the telecom bands, e.g. 1550 nm, we focus on the theoretical design of a programmable mode converter (PMC) device utilizing the contrast in refractive index $n$. This approach is fundamentally different from the conventional GST-based waveguide device, which relies on the telecom band lasers for both programming (finite $k$ values of both $a$- and $c$-GST) and signal detections ($\Delta k$ between $a$- and $c$-GST). Hence, the scalability of GST photonic array is seriously limited by the high optical loss induced by the ample free carriers of its crystalline phase. However, with the $Sb_2Se_3$-based PMC design, optical loss is no longer a major issue. The low-loss feature together with sizable $\Delta n$ in the telecom bands of $Sb_2Se_3$ is expected to largely extend the array size for photonic computing. Nevertheless, the finite-loss feature of both $a$- and $o$-$Sb_2Se_3$ in the visible-light range is still a compulsory ingredient for triggering the reversible phase transition.

Fig. 3a shows a 3D schematic of the proposed device for programmable mode converting from the incident fundamental transverse electric ($TE_0$) mode to the first-order transverse electric ($TE_1$) mode. We used an air-cladding 70-nm-thick SOI strip waveguide that was shown to have a very low propagation loss[71] as the basic platform. Then we considered a shallow-etched parallelogram slot filled with a 25-nm-thick $Sb_2Se_3$ patch and a 25-nm-thick $SiO_2$ capping layer. This ultrathin SOI waveguide has a weak optical confinement to enhance light interaction with the PCM layer. The eigenmode profiles of the $TE_0$ and $TE_1$ modes at 1550 nm overlapping with the cross-section of waveguide are shown in the Fig. 3b. For the finite-difference time-domain (FDTD) simulations, we set the *k* values as 0, and used *n* = 3.28 and 4.05 for amorphous and crystalline $Sb_2Se_3$ at 1550 nm (Fig. 2c). The difference in the refractive index between crystalline (amorphous) $Sb_2Se_3$ and silicon is 0.58 (−0.19), respectively.

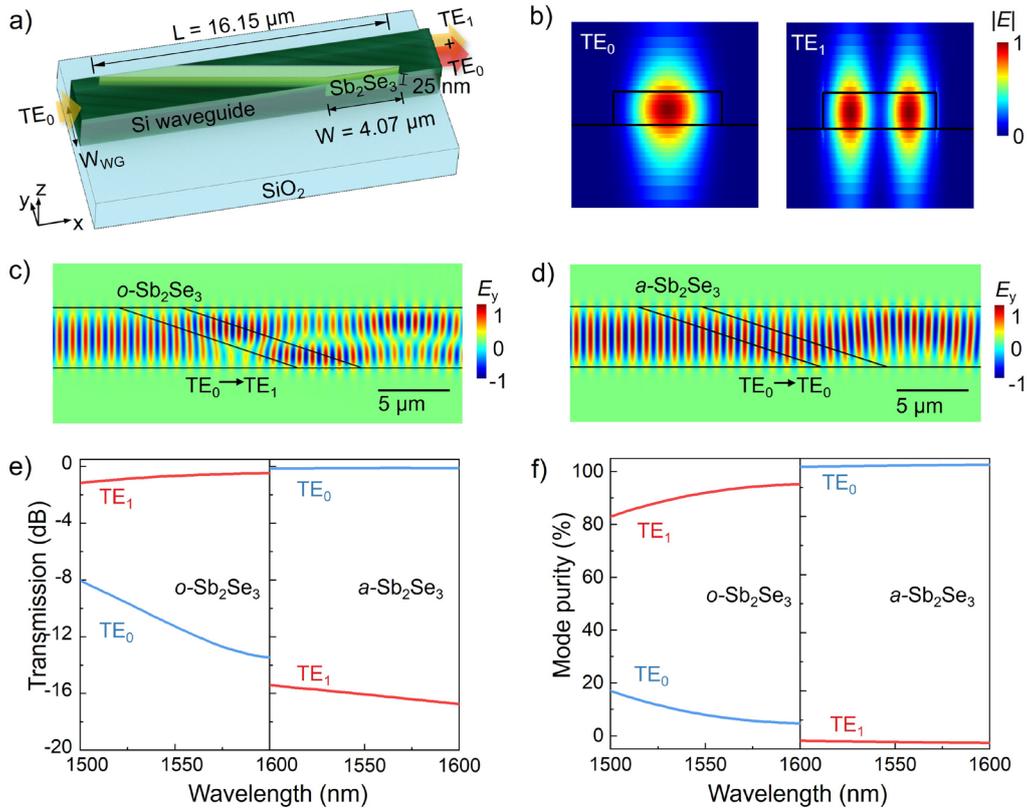

**Fig. 3 Schematic and the performance of the waveguide mode converter.** a) 3D schematic of $Sb_2Se_3$-based waveguide mode converter. b) Modal profiles of the $TE_0$ and $TE_1$ mode. c-d) Electric modal profiles ($E_y$ component) of the waveguides with c) *o*-$Sb_2Se_3$, and d) *a*-$Sb_2Se_3$. e) Transmission spectra of devices with *o*-$Sb_2Se_3$ and *a*-$Sb_2Se_3$ for $TE_0$ and $TE_1$ modes. f) Mode purity spectra of devices with *o*-$Sb_2Se_3$ and *a*-$Sb_2Se_3$ for $TE_0$ and $TE_1$ modes.

The presence of a $Sb_2Se_3$ patch can induce a dielectric perturbation ($\Delta\varepsilon$) in the silicon waveguide for achieving mode conversion[72]. The modal coupling coefficient ($K_{ij}$), i.e., the coupling strength between the *i*-th and *j*-th ordered modes in the waveguide, is expressed as[73]:

$$K_{ij}(x,y,z) = \frac{\omega\varepsilon_0}{4} \int \boldsymbol{E}_p^*(x,y) \cdot \Delta\varepsilon(x,y,z) \cdot \boldsymbol{E}_q(x,y) dx dy \quad (3)$$

A larger $K_{ij}$ can be obtained using o-$Sb_2Se_3$ than that of a-$Sb_2Se_3$ due to a larger $\Delta\varepsilon$. Moreover, for efficient and selective mode converting using o-$Sb_2Se_3$, phase matching condition of $\Delta\beta_{o\text{-}Sb2Se3} = \beta_i - \beta_j - \beta_{o\text{-}Sb2Se3} = 0$ should be satisfied, while $\Delta\beta_{a\text{-}Sb2Se3} \neq 0$ to prohibit mode converting for the device with a-$Sb_2Se_3$, where $\beta_i$ and $\beta_j$ are the propagation constants of the i-th and j-th ordered modes. The $\beta_{o\text{-}Sb2Se3}$ and $\beta_{a\text{-}Sb2Se3}$ (= $2\pi j/L$) are the compensating propagation constants provided by the dielectric perturbation using o-$Sb_2Se_3$ and a-$Sb_2Se_3$ patches, respectively. Here, we focus on the mode conversion between $TE_0$ and $TE_1$ modes, such that subscripts $i = 0$ and $j = 1$, $\beta_{o\text{-}Sb2Se3} = 2\pi/L$, and $L = 2\pi/(\beta_0 - \beta_1)$. We set the width of the silicon waveguide ($W_{WG}$) as 2.25 μm to only support $TE_0$ and $TE_1$ modes. The calculated effective refractive index values of $TE_0$ and $TE_1$ modes were $n_{eff0} = 1.784$ and $n_{eff1} = 1.688$, respectively. Therefore, $L$ was derived as 16.15 μm. As discussed in Ref.[72], $W < L/2$ should be satisfied to maintain $K_{01}(z) \neq 0$ through the entire conversion process. Next, we set $W$ to be equal or close to $L/4$ and optimized the geometric parameters $W$ and $L$ to obtain a high transmittance for $TE_1$ mode with efficient mode conversion from the incident $TE_0$ mode through inter-modal coupling. The optical transmittance is defined as: $T = 10 \cdot \log_{10}[P_{out}/P_{in}]$, where $P_{in}$ and $P_{out}$ are the input power and transmitted power of $TE_0$ ($TE_1$) mode, respectively. The mode purity is defined as: $\eta_{TE0(TE1)} = P_{TE0(TE1)}/(P_{TE0}+P_{TE1})$.

Fig. 3c shows the simulated electric field ($E_y$) evolution for the device with $W = 4.07$ μm and $L = 16.15$ μm. It can be clearly observed that the input $TE_0$ mode is converted to the $TE_1$ mode after propagating through the o-$Sb_2Se_3$ patch. However, as shown in Fig. 3d, the a-$Sb_2Se_3$ patch only induces limited perturbation to the propagation of the input $TE_0$ mode without obvious conversion to the $TE_1$ mode due to unsatisfied phase matching condition $\Delta\beta_{a\text{-}Sb2Se3} \neq 0$. Therefore, this waveguide device achieves selective mode conversion upon phase transition of $Sb_2Se_3$. As shown in Fig. 3e and Fig. 3f, the transmittance (mode purity) values for $TE_0$ and $TE_1$ modes of the o-$Sb_2Se_3$ device are −11.33 dB and −0.65 dB (7.88% and 92.12%) at 1550 nm, respectively. The transmittances (mode purity) values for $TE_0$ and $TE_1$ modes of the a-$Sb_2Se_3$ device are −0.11 dB and −16.08 dB (97.53% and 2.47%) at 1550 nm, respectively. We also achieved a broadband operating window that ranges from 1500 nm to 1600 nm, which is advantageous for parallel photonic computing based on the WDM scheme[31,32]. In contrast to the sub-100 nm metasurface-based mode converter[34], the feature size of our device is larger than 2 μm, largely reducing the fabrication complexity.

To make the optical detection process more straightforward, we added an asymmetric directional coupler to convert $TE_1$ mode from the bus waveguide to $TE_0$ mode and output it through a second port. As shown in Fig. 4a (right side), the coupler contains three critical parameters, namely, the width of the bus waveguide $W_{bus}$, the width of the side coupled dropping waveguide $W_{drop}$ and the coupling length $L_c$. To satisfy the phase-matching condition[74], the effective refractive index of the $TE_1$ mode in the bus waveguide must be equivalent to that of the $TE_0$ mode in the side coupled dropping waveguide, namely, $n_{eff1}(W_{bus}) = n_{eff0}(W_{drop})$. However, the unconverted $TE_0$ mode would propagate along the bus waveguide directly to the output port 1 without coupling to the output port 2 due to phase mismatching.

Ideally, the input TE$_0$ mode is only detected at output port 1 for the a-Sb$_2$Se$_3$ device. For the o-Sb$_2$Se$_3$ device, the input TE$_0$ mode is firstly converted to the TE$_1$ mode by our mode converter and then coupled back to the TE$_0$ mode by the asymmetric directional coupler, and finally detected only at the output port 2. However, neither the conversion from the TE$_0$ mode to the TE$_1$ mode through the o-Sb$_2$Se$_3$ patch is 100% nor all the input TE$_0$ mode would fully pass through the a-Sb$_2$Se$_3$ patch without conversion. After optimization, we obtained $W_{drop}$ = 0.86 μm and $W_{bus}$ = 1.8 μm with calculated $n_{eff1}(W_{bus}) = n_{eff0}(W_{drop})$ = 1.63, and $L_c$ = 27.86 μm for the asymmetric directional coupler. At 1550 nm, the FDTD simulations show that the detected mode purity of the TE$_0$ mode to be 7.58% at output port 1 and 92.42% at output port 2 for the o-Sb$_2$Se$_3$ device, and 99.79% at output port 1 and 0.21% at output port 2 for the a-Sb$_2$Se$_3$ device. We can obtain a mode contrast $\Gamma = \eta_{port1} - \eta_{port2}$ = −0.85 for the o-Sb$_2$Se$_3$ device and 0.99 for the a-Sb$_2$Se$_3$ device.

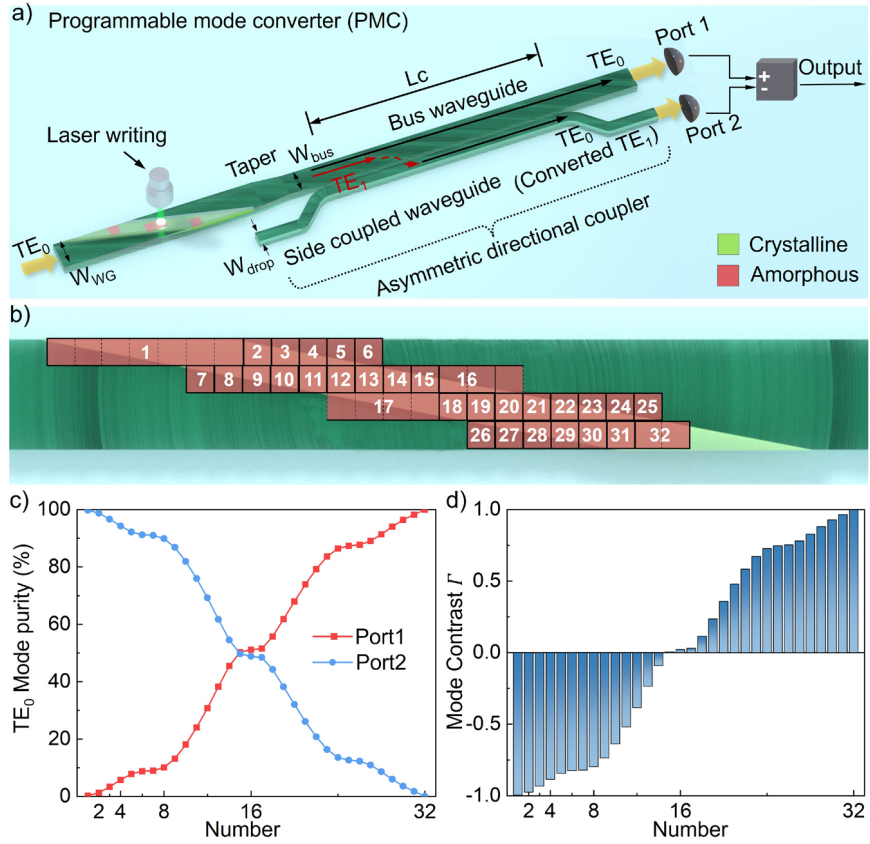

**Fig. 4 Programmable mode converter (PMC) device and its performance.** a) 3D illustration of the PMC device for mode-selective coupling. b) Top view of labeled 32 segments of a Sb$_2$Se$_3$ patch. c) The mode purity and d) the mode contrast values detected at port1 and port2 with a step-by-step amorphization of the Sb$_2$Se$_3$ patch.

As shown in Fig. 2c, the *k* values of Sb$_2$Se$_3$ in the visible light range are non-negligible, thus it is feasible to induce phase transition using visible light lasers. It has been demonstrated that a small area of 0.75 × 0.75 μm$^2$ on Sb$_2$Se$_3$ thin films could be switched from the crystalline state to the amorphous state using lasers of 638 nm, and a feature pattern size of ~0.3 μm amorphized strips was achieved on Sb$_2$Se$_3$ thin films using lasers of 405 nm[57,58]. Here, we considered a patterning feature size of 0.575 μm, and divided an entire $o$-$Sb_2Se_3$ patch into multiple programming areas to achieve 32 distinct levels. The detailed segments are labeled in Fig. 4b and each pixel has a floor area of 0.575 × 0.575 μm$^2$. For a full crystalline $o$-$Sb_2Se_3$ device, the $TE_0$ mode purity cannot reach ~100% at output port 2, but such high value can be obtained when the device is in a partial amorphous/crystalline configuration. More specifically, when the first 7 pixels are in the amorphous phase and the rest 37 pixels are in the crystalline $o$-phase, the $TE_0$ mode purity reaches 99.74% at output port 2. This configuration corresponds to the state 1. To gradually decrease the $TE_0$ mode purity at output port 2 (and thereby an increase in the $TE_0$ mode purity at output port 1), the $Sb_2Se_3$ pixels need to be amorphized step by step. However, the change in $TE_0$ mode purity does not scale linearly with the increase in number of amorphous pixels. For the state 1, state 16, state 17 and state 32, multiple pixels are need to be amorphized together to result in a sizable change in $TE_0$ impurity. As shown in Fig. 4c, the mode purity can be sequentially tuned from 0.25% to 99.98% for output port 1 and from 99.74% to 0.02% for output port 2. As a result, mode contrast $\Gamma$ is between −0.995 and 0.999 as shown in Fig. 4d. It is worth noting that a smaller pattern size could in principle result in more tunable number of levels. Mathematically, the whole PMC device represents a multiply accumulate operation of ($G \cdot X + b$), which is the foundation of the matrix-vector multiplication (MVM) operations. Here, $X$ is the input light power, $G$ can be programmed by setting $\Gamma$ in the mode converter, and $b$ is the bias parameter representing the accumulation variable. When the encoded input $TE_0$ mode propagates through PMC, due to the satisfied phase matching condition, it will be converted to $TE_1$ mode. The intensity of the converted $TE_1$ mode is determined by the fraction of amorphous/crystalline state of $Sb_2Se_3$, namely, the mode contrast value ranging from −0.995 and 0.999. In a multimode photonic array (as shown in Fig. 5b), the encoded input laser beams are sent to the waveguides as an input vector. After converted by PMC, the intensities of the output $TE_0$ and $TE_1$ modes (the $TE_1$ mode is converted to $TE_0$ mode again via the side coupler) are accumulated at output port 1 and port 2 through the same column in the array, and the results of the multiply accumulate operation can be detected using a balanced photodetector.

Overall, our $Sb_2Se_3$-based programmable mode converter device is capable of achieving a balance among the footprint, insertion loss, reliability, and integrating density, which may render it a suitable candidate for photonic computing applications. A detailed comparison between our device and other PCM-based photonic devices is provided in Table 2. We also briefly discuss the fabrication feasibility of the $Sb_2Se_3$-based PMC device. The base SOI waveguides can be produced by using deep ultraviolet (DUV) based multi-project wafer (MPM) services, such as those offered by the common 180-nm Complementary Metal Oxide Semiconductor (CMOS) technology node in the silicon photonic foundries. The back-end of line (BEOL) processing can be completed via photolithography etching and thin film deposition, as shown in Fig. S7. The programming of the $Sb_2Se_3$ PMC device for phase transition can be done using a light and digital micromirror setup reported in Ref.[75]. A possible programming process of the $Sb_2Se_3$ PMC device is depicted in Fig. S8.

| Type of Device | Year | PCM | Feature size (μm) | Insertion Loss (dB) | Number of levels | Operation bandwidth | Programming mode |
|---|---|---|---|---|---|---|---|
| MRR[Ref.53] | 2025 | $Sb_2Se_3$ | 13 | 6 | / | Narrowband | Electrical |
| MRR[Ref.56] | 2024 | $Sb_2Se_3$ | ~40 | 0.459 | 36 | Narrowband | Electrical |
| MZI[Ref.56] | 2024 | $Sb_2Se_3$ | >100 | ~2.6 | / | Broadband | Electrical |
| MZI[Ref.55] | 2023 | $Sb_2Se_3$ | >100 | ~2.5 | >32 | Broadband | Electrical |
| MZI[Ref.54] | 2022 | $Sb_2Se_3$ | >100 | ~5 | 9 | Broadband | Electrical |
| **This work** | **2025** | $\mathbf{Sb_2Se_3}$ | **16.15** | **0.65** | **32** | **Broadband** | **Optical** |

**Table 2.** Comparisons of our PMC design with other state-of-the-art MRR and MZI photonic devices based on $Sb_2Se_3$, including the feature size, insertion loss, number of levels, operational bandwidth, and the type of programming mode. "/" means not mentioned.

**Scalability of the PMC array**

For practical applications, it is important to have large array size. Next, we evaluate the scalability of the $Sb_2Se_3$-based PMC array. We also considered the standard crossbar array using GST for comparison[33]. Fig. 5a shows the schematic of a single-mode photonic array using the crossbar PCM devices as the computing core to perform MVM operations in the form of M· X = Y. The input vector values from $X_1$ to $X_m$ are encoded as the input light power values at different wavelength, and the matrix weights M are presented by the specific phase configuration of GST devices in the crossbar array. The output vector values from $Y_1$ to $Y_m$ are the resulting light power values of the dot product between the input light power values and the absorption ratio of the devices in each column in the matrix M. Another scheme of the photonic tensor core employs mode division methods, and the typical multimode photonic array architecture is shown in Fig. 5b. With wavelength division multiplexers (MUXs) and wavelength division demultiplexers (DEMUXs), a series of input light can be encoded on different spatial modes, then can be sent into the device array simultaneously. This scheme is suitable for our $Sb_2Se_3$ PMC devices.

Fig. 5c shows a representative GST waveguide cell (2×2 μm$^2$) for the single-mode photonic array[33] and a $Sb_2Se_3$ PMC device for the multimode photonic array. For crystalline GST, the high extinction coefficient at 1550 nm leads to strong absorption of the passing optical signal, while in the amorphous phase, the absorption is much reduced due to the lower extinction coefficient. We simulated the insertion loss, which is defined as $D = 10·\log_{10}[P_{out}/P_{in}]$, of the GST crossbar device. In practical photonic array employing reconfigurable PCMs, the maximum array size is typically limited by the large optical loss induced by crystalline GST waveguide cells with high amounts of free carriers. According to our FDTD simulations, the 2×2 μm$^2$ GST cell shows a very high loss of −14.97 dB in the fully crystalline phase, but only −1.71 dB in the fully amorphous state. Based on these insertion loss values, we estimated the optical loss of the system as a function of matrix size (m) of the photonic arrays, as shown in Fig. 5d. Here, the scalability of the photonic matrix is considered to be limited by the insertion loss (D) of the PCM devices and the light splitting ratio (1/m) of the waveguide array[34], leading to the following equation: $D+10\times\log_{10}(1/m^2)$. By setting an empirical total loss of −43 dB as the detectable threshold, our simple estimate suggests that the maximum size of the GST waveguide array is limited to 26×26 (Fig. 5d blue dots), when all the waveguide cells are programmed to the fully crystalline phase. Such array size could reach 116×116 for the fully amorphous state. In experiments, the maximum matrix size was found to be ~32×32[33], although the specific crystalline/amorphous configuration in each waveguide cell was not provided. Nevertheless, our estimate with the highest-possible optical loss provides a reasonable estimate of the waveguide array size. We note that our $Sb_2Se_3$ PMC design shows a much lower insertion loss, reaching only −0.65 dB per device. By considering −43 dB as the threshold, the estimated $m$ is calculated to be ~131. This suggests that our $Sb_2Se_3$ PMC array could in principle support a matrix size of >128×128 (Fig. 5d red dots) for more efficient MVM operations.

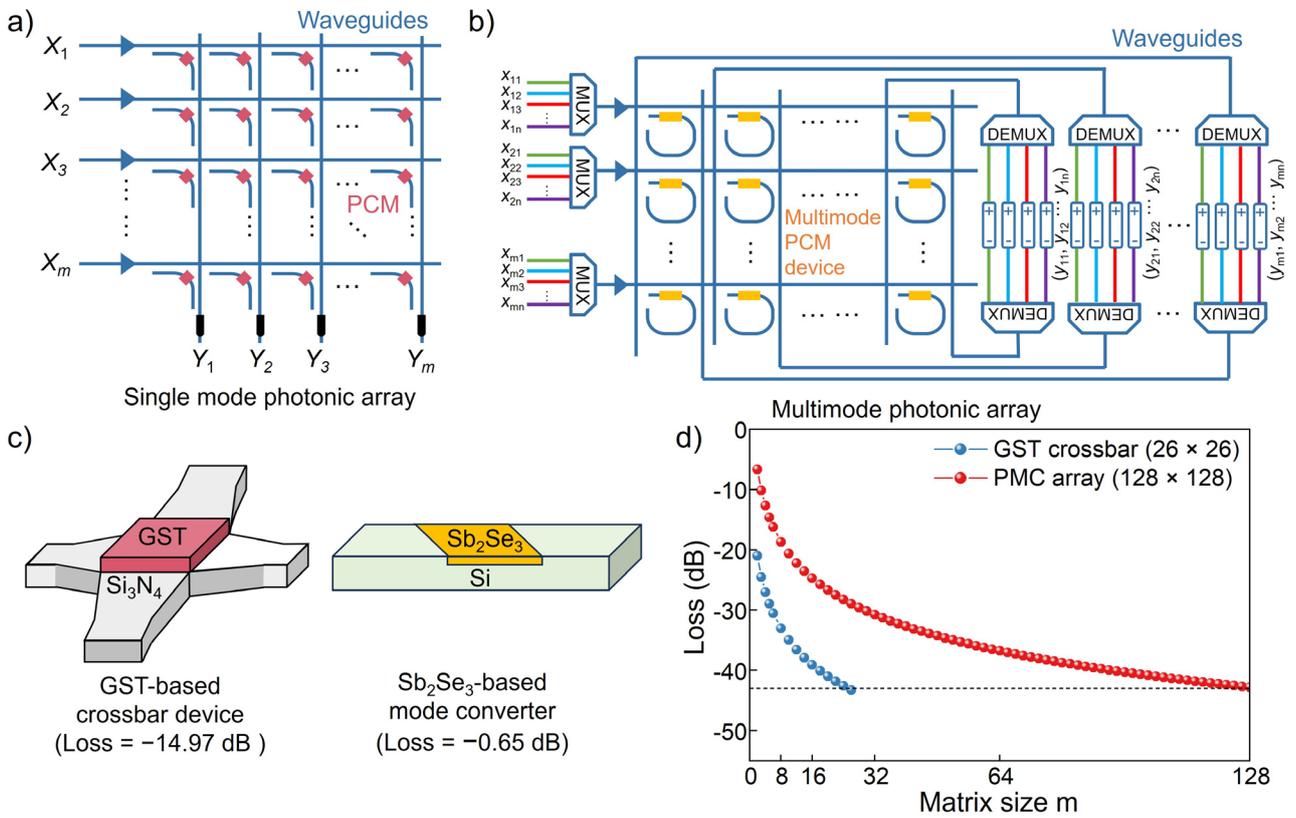

**Fig. 5 Schematics of waveguide arrays and the estimated matrix size.** a) Schematic of a single-mode photonic array. b) Schematic of a multimode photonic array with wavelength division multiplexers (MUXs) and wavelength division demultiplexers (DEMUXs). c) Illustrations of the crossbar device and PMC device. The corresponding insertion loss at wavelength of 1550 nm obtained by FDTD simulations are also shown. d) The estimated optical loss as a function of matrix size (m) of the two photonic arrays. The dashed line marks the optical loss of −43 dB.

## Image convolution and recognition using the PMC array

Finally, we carried out a couple of numerical simulations using the $Sb_2Se_3$ PMC array. Fig. 6a illustrates the mechanism of image convolution, a fundamental operation in image processing and computer vision. A 3×3 kernel matrix is placed on the top of the input image, and gradually slide from left to right and top to bottom with a fixed interval of pixel. For each step, the kernel is multiplied with the corresponding pixels on the input image, and the results are summed into a single value to represent a new pixel in the output image. In the PMC array, the kernel weights are represented by the mode contrast values. The values of pixels in the original image are encoded onto light powers of nine separated wavelength channels after applying intensity modulation. The output image corresponds to a time series of patch-kernel MVM operations. With summation of incoherent light signals output from multiple mode converting devices, $Y = \sum(M_i \cdot X_i)$ can be performed in the photonic tensor core. Besides, the balanced photodetection scheme[76] generates both positive and negative signals, such that 2-quadrant multiplication can be implemented in the photonic tensor core without the need of an additional offset, which otherwise requires additional data post-processing.

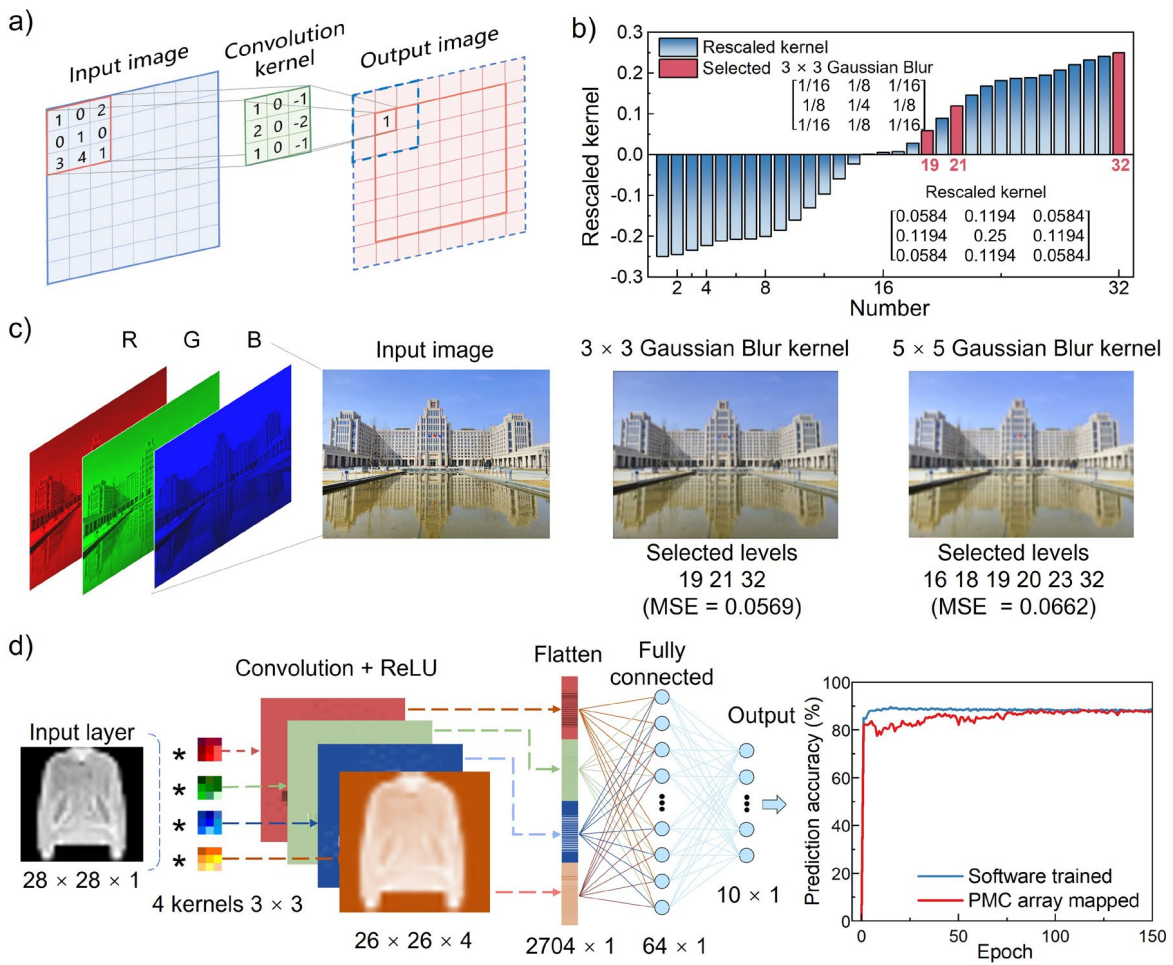

**Fig. 6 Convolutional image processing and convolutional neural network simulations using the PMC array.** a) Convolutional processing of an input image. b) Selected mode contrast levels for a 3×3 Gaussian blur kernel. c) The input image and processed images after convolving with the 3×3 and 5×5 Gaussian blur kernel. The input image showing the Hanying building in the Xi'an Jiaotong University was photographed by the authors. The selected levels as well as the mean square error (MSE) are shown below the image. d) The CNN framework and the prediction accuracy in performing the MINIST fashion dataset recognition task.

As shown in Fig. 6b, we rescaled the range of 32 mode contrast values from [−0.995, 0.999] to [−0.25, 0.25], and mapped the element values into a 3×3 Gaussian blur kernel with the closest weights highlighted by the red color bars. The obtained selected states for this kernel are 19, 21 and 32, forming a rescaled kernel matrix based on our PMC array, see the two matrices in Fig. 6b. An original input image with a size of 256×192 and three RGB color channels were convolved with the rescaled 3×3 Gaussian blur kernel. The output image is shown in the middle panel of Fig. 6c. A clear blurring effect can be observed in the output image. The mean square error (MSE) of the computing error, 0.0569, between the results of the standard kernel and PMC array kernel is quite low, indicating a comparable processing accuracy. We also modeled a 5×5 Gaussian blur rescaled kernel with 6 selected levels, which resulted in a more blurred output image. The detailed standard kernel values and rescaled kernel values are displayed in Fig. S9. Since our PMC array could support a matrix size of >128×128, many more kernel values could be stored in the array simultaneously and the supported kernel size can be scaled up to 7×7, 9×9 or even higher. In Fig. S10, we showed the processed images using a 7×7 Laplacian edge detection kernel and a 9×9 Gaussian Laplacian of Gaussian kernel. The obtained MSE values are also quite low, 0.0687 and 0.0444.

In addition, we explored the application of PMC array in convolutional neural network (CNN) for image recognition tasks. Using the MNIST fashion dataset[77] as input, Fig. 6d illustrates the CNN architecture and the prediction accuracy of PMC array by kernel mapping compared with software training. First, fashion product images with a pixel size of 28 × 28 were convolved with four 3 × 3 kernels, generating four 26 × 26 activation maps. Then, the ReLu activation function[78] was implemented, and the output images were finally classified into 10 groups after data flattening and processing by fully connected layers. The resulting prediction accuracy of software training achieves 88.6% after 150 training epochs, and the four optimized kernel matrix weights were obtained through software training. Next, the square mapping error (SME) was introduced as the mapping function between the 32 mode contrast levels of the PMC array and the optimal kernel matrix weights. SME is defined as the total square differences between the standard weight $A$ and the mapped weight $B$[79]:

$$SME = min\left\{\sum (\lambda B - A)^2\right\} \quad (4)$$

By optimizing the value of mapping coefficient $\lambda$, the weight matrix in CNN was reset, and the recognition accuracy of PMC array was re-evaluated. The prediction accuracy of mapped network was gradually improved with the increase in the number of training epochs, eventually reaching 87.6% at 150 epochs. Such accuracy is comparable to the software trained result. We also performed recognition task of the MNIST handwritten digits dataset[80], the predicted recognition accuracy could reach 97.8% after 150 training epochs (Fig. S11). Overall, our CNN simulations predicted a competitive level of accuracy in these recognition tasks as compared to literature data (see Table S1).

Before closing, we discuss some limitations and potential device non-idealities of $Sb_2Se_3$ based devices. It is known that the cycling endurance of $Sb_2Se_3$ is typically limited to $10^3$–$10^5$ cycles, while that of GST can reach $10^9$–$10^{12}$ cycles. Besides, crystallization of $Sb_2Se_3$ needs ms-level pulses, but 10 ns pulse is already sufficient to crystallize GST. These limitations of $Sb_2Se_3$ can be attributed to the more complex crystalline structure of orthorhombic $Sb_2Se_3$, which could potentially result in a more difficult crystallization pathway. As seen in Fig. 1a, the orthorhombic phase of $Sb_2Se_3$ has a zigzag-shaped structural gaps with Se⋯Se long-term atomic contacts, and short and strong Sb–Se bonds with strong bond distortions inside each atomic slab. This structural feature stems from the stronger $sp^3$ mixing induced by the abundant Se atoms[81]. In comparisons with cubic rocksalt $Sb_2Te_3$ and GST, where all atoms form octahedral bonds with no discontinuous structural gaps, a high fraction of Sb and Se atoms form non-octahedral motifs and a high density of zigzag-shaped structural gaps is present. Upon rapid and repeated phase transition cycles between amorphous and crystalline phase, the more complex crystalline structure with strong bond distortions may increase the possibility of forming excess structural defects in $Sb_2Se_3$. The accumulated defects could result in unwanted cycle-to-cycle and device-to-device variabilities. In addition, resistance drift issue is known to be present for amorphous $Sb_2Se_3$ as well[82], and the widening of band gap with time may also alter its optical properties in the visible-light range for phase switching. Better materials confinement, e.g. thick capping layers up to sub-μm, and additional thermal annealing after phase switching may help improve the cycling endurance and programming consistency of $Sb_2Se_3$ based photonic devices. When implementing the PMC devices in a high-density array, it is also important to monitor the potential thermal crosstalk between adjacent waveguide cells. A system-level validation algorithm needs to be developed as well. For practical applications, the system-level non-idealities could also affect the scaling of the PMC array. For instance, additional insertion losses may arise from, e.g., directional couplers, which scale linearly with the matrix size. Note that our estimate of array size is based on the maximum loss of fully crystalline devices. In practice, different devices are programmed to a mixture of crystalline and amorphous states, which could result in lower optical losses. Further experimental and engineering efforts are much needed to verify the predicted performances of $Sb_2Se_3$ in a practical photonic array, before this low-loss material can be used for practical photonic computing applications.

**Discussion**

Finally, we make a comprehensive comparison between low-loss PCMs and conventional MVB-type PCMs for photonic applications, ranging from the basic bonding mechanism to the design of device structure, as summarized in Fig. 7. For both types of PCMs, their density values are typically reduced upon amorphization. The crystalline structures of low-loss PCMs, e.g., $Sb_2Se_3$, $Sb_2S_3$, and GSST alloys, are more distorted as compared to conventional MVB-type PCMs, e.g., GeTe, GST, and Ag-In-Sb-Te alloys, because of the incorporation of Se or S atoms. In comparison with Te, these lighter chalcogen elements show a stronger $sp^3$ mixing tendency, resulting in only partly aligned $p$ orbitals. Regarding the amorphous phase of all these materials, their $p$ orbitals are in general misaligned due to the absence of long-range order. Here, we take GeTe as an example for conventional PCM. The rock-salt phase of GeTe shows a perfect alignment of *p* orbitals, and this MVB mechanism leads to high probabilities of inter-band transitions, as reflected by the calculated TDM. Upon amorphization, the alignment of *p* orbitals is broken, resulting in a large contrast window in both *n* and *k* across the visible-light and telecom wavelength range. The sizable Δ*k* at 1550 nm is typically utilized to detect different optical states for conventional PCM based single-mode photonic device. For $Sb_2Se_3$, both Δ*n* and Δ*k* values are reduced because of the absence of MVB in both crystalline and amorphous phases. Considering the larger optical band gaps, the *k* values of both a-$Sb_2Se_3$ and c-$Sb_2Se_3$ are close to zero at the telecom wavelength range. Although the Δ*n* value at 1550 nm of $Sb_2Se_3$ is smaller than that of GeTe, it is already sufficient to detect multiple optical states via the mode converter approach discussed above.

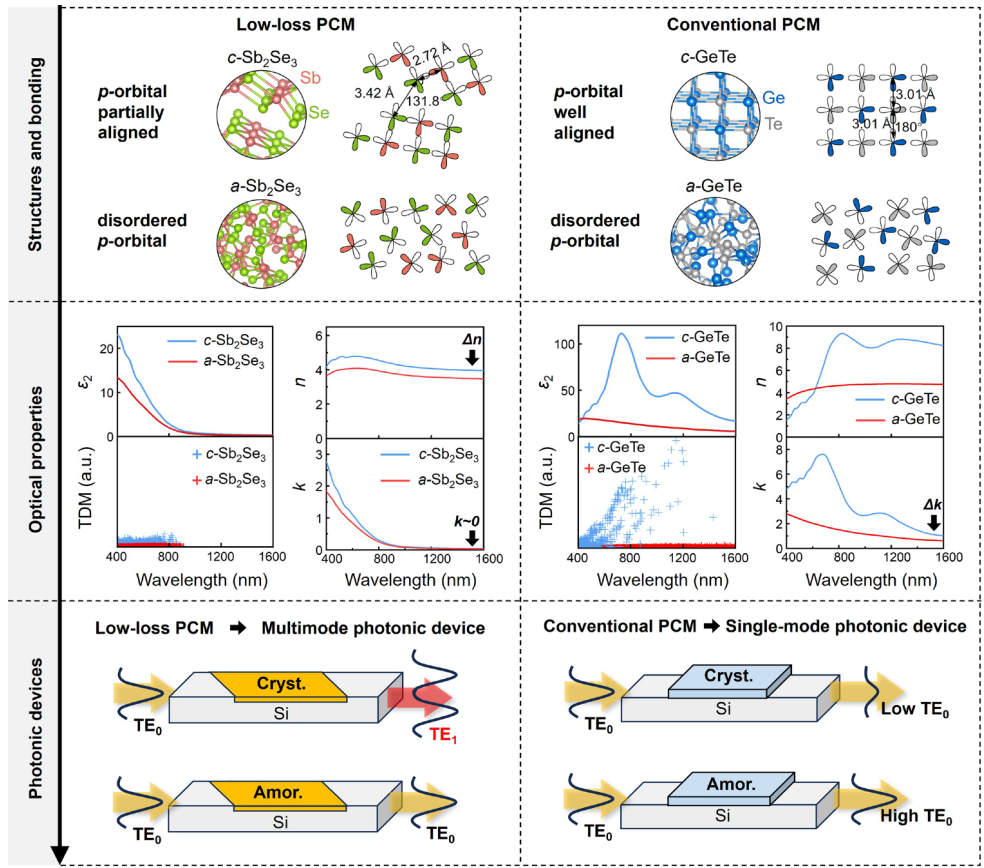

**Fig. 7. Comparisons between low-loss PCM and conventional MVB-type PCM.** The difference in bonding characteristics leads to a distinct programming approach for all-optical photonic applications.

In summary, we employed multiscale simulations to investigate bonding origin of low-loss phase-change alloy $Sb_2Se_3$ and designed a programmable mode converter waveguide device for photonic neuromorphic computing applications. We designed a mode converting device based on $Sb_2Se_3$, which can be programmed to 32 distinct states in a micro-sized device via direct laser writing. With an insertion loss as low as 0.65 dB per device, this PMC design could potentially support the scalability of photonic tensor cores to matrix sizes >128×128. Numerical simulations showed that this $Sb_2Se_3$-based multimode photonic tensor core could be exploited for convolutional image processing and recognition applications with high accuracies. The current all-optical setup could enable a superior spatial selectivity (~0.33 µm² per pixel) for accurate programming using focused laser pulses. However, it also requests an external beam-steering setup to induce phase transition in the $Sb_2Se_3$ patch, which may slow down the overall operation speed. In parallel to all-optical computing, embedded microheaters are under active development, which could allow simultaneous electrical programming and optical computing[32]. The thermal field generated by a microheater device covers a much larger area[83], e.g., ~35 µm². Using segmented and doped Si microheaters[83], the switchable area of PCM can be reduced to less than 3 µm². To avoid potential thermal crosstalk problem, a spatial separation of 10 µm between adjacent waveguide cells[33] could be necessary for high-density integration. Overall, this work provides an atomic-scale understanding of the low-loss mechanism of $Sb_2Se_3$ and establishes a pathway for its integration into scalable photonic array. We anticipate that our theoretical efforts will stimulate further experimental exploration of low-loss PCMs for photonic neuromorphic computing applications.

## Methods
### Ab initio calculations
We carried out AIMD simulations by using the Vienna Ab initio Simulation Package (VASP)[84]. The projector augmented-wave (PAW) pseudopotentials[85], the Perdew–Burke–Ernzerhof (PBE) functional[86] and the Grimme's DFT-D3 method for van der Waals interactions[87] were used. The energy cutoff was set as 350 eV. The time step was set as 2 fs, the NVT ensemble and Nose-Hoover thermostat were used for controlling the temperature. The amorphous models were firstly randomized at 3000 K for 15 ps, then quenched to 1200 K, and kept at this temperature for 30 ps. Then, the models were cooled down to 300 K with a rate of 25 K/ps, and were kept at 300 K for 30 ps to collect structural data. The models were finally cooled to 0 K for structure relaxation. The cell edge was adjusted during the simulation to ensure that the internal pressure remains below 3 kbar. Only $\Gamma$ point was used to sample the Brillouin zone of amorphous models. The $K$-point meshes used for the relaxations of o-, m-, r-, and s-r-$Sb_2Se_3$ were 3 × 9 × 3, 2 × 3 × 6, 7 × 7 × 1, and 7 × 7 × 1, respectively. To obtain more accurate electronic structure and optical profiles, we performed VASP calculations using the HSE06 functional[70]. The $k$-point mesh numbers were increased 2 times for optical calculations. For the ET and ES calculations, the Quantum Espresso (QE) code[88] was used to produce all-electron density for the models by both PAW and norm-conserving potentials together with PBE and DFT-D3 correction. The energy cutoff for wavefunctions is 80 Ry and for charge density is 320 Ry, respectively. Then, the QE wavefunctions were directly processed by Critic2 code[89] to determine the atomic basins, and Domain Overlap Matrices (DOM) were computed to get the delocalization (DIs) and localization indices (LIs). The ET value was obtained by firstly computing the difference between the total electrons in the atomic basin and the free reference atom, then normalizing this difference by the formal oxidation state of the atom. ES values were twice of the DI values between adjacent atoms.

## FDTD simulations

The PMC waveguide devices were modeled using the Lumerical FDTD Solutions code[90]. The refractive indices data for Si, $Si_3N_4$ and $SiO_2$ were taken from the built-in material database. The experimentally measured refractive indices data for GST and $Sb_2Se_3$ were taken from Ref.[52]. Perfectly matched layer (PML) boundary condition was applied for all the boundaries. The fundamental transverse electronic ($TE_0$) mode was injected into the waveguide with a wavelength range between 1500 nm and 1600 nm. The $TE_0$ mode is placed at x = −8 μm and propagates along the x direction. The $Sb_2Se_3$ patch and the capping $SiO_2$ layer are positioned with the left edges at x = 0 μm. A trapezoidal taper with a total length of 5 μm is located at x = 0 μm to connect the waveguide with mode converter and the bus waveguide. The distance between the bus waveguide and side coupled dropping waveguide is optimized as 0.18 μm. Mode expansion monitors were used for extracting transmission spectra of the $TE_0$ and $TE_1$ modes. The electric field profiles $|E_y|$ are also calculated by the mode monitors. The conformal variant 0 mesh refinement method is applied. Grid sizes were set fine enough to obtain converged simulation results.

## Image convolution and CNN simulations

The kernel rescale process and image convolution results were simulated by MATLAB. The construction, training, and weight mapping using SME in the CNN are realized by Python PyTorch package.

## Data Availability

The data that support the findings of this study will be available from in [CAID Repository] at [https://caid.xjtu.edu.cn/info/1003/2051.htm] upon journal publication.

## Code Availability

The authors declare that the applied software supporting the findings of this study are commercially available in the VASP software package https://www.vasp.at and Lumerical FDTD Solutions https://www.ansys.com/products/photonics/fdtd.

## Competing interests

The authors declare no competing interests.

## Ethical approval

This study does not involve human participants, animal subjects, or sensitive personal data.

## Acknowledgements

The work is supported by the National Key Research and Development Program of China (2023YFB4404500). W.Zhang thanks the support of the National Natural Science Foundation of China (62374131). W.Zhou thanks the support of the National Natural Science Foundation of China (62405242). The authors thank the supported by the 111 Plan (B25007). The authors acknowledge the computational resources provided by the HPC platform of Xi'an Jiaotong

University and the Computing Center in Xi'an. The authors acknowledge the International Joint Laboratory for Micro/Nano Manufacturing and Measurement Technologies of Xi'an Jiaotong University.

**Supporting Information**

Supporting Information is available.

# Supplementary information

# for

## *Low-loss phase-change material based programmable mode converter for photonic computing*

Xueyang Shen[1], Ruixuan Chu[1], Ding Xu[1], Yuan Gao[1], Wen Zhou[1*], Wei Zhang[1*]

[1]Center for Alloy Innovation and Design (CAID), State Key Laboratory for Mechanical Behavior of Materials, Xi'an Jiaotong University, Xi'an, 710049, China.

*Emails: wen.zhou@xjtu.edu.cn, wzhang0@mail.xjtu.edu.cn

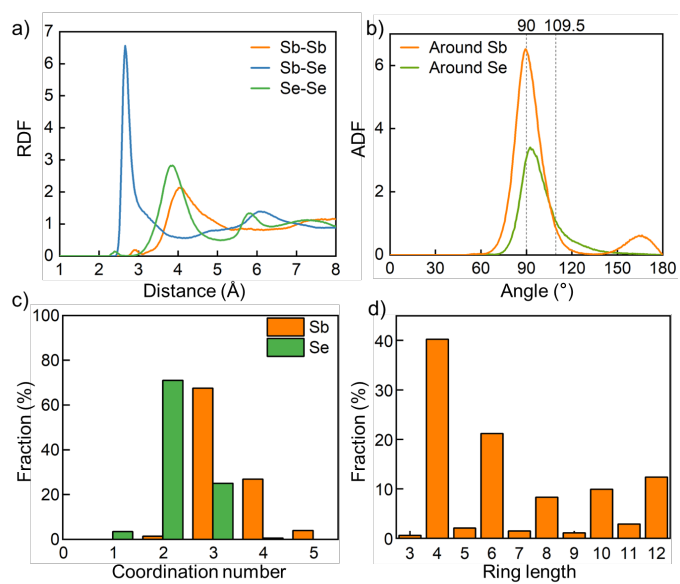

Fig. S1. AIMD simulations of amorphous $Sb_2Se_3$. a) The radial distribution function (RDF), b) angular distribution function (ADF), c) coordination number, and d) ring length distribution of *a*-$Sb_2Se_3$ models. The cutoff values of 3.10 Å for Sb–Sb, 3.00 Å for Sb–Se, and 2.70 Å for Se–Se contacts were used for structural analysis.

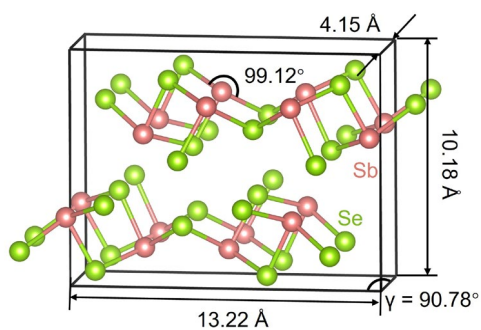

Fig. S2. The atomic structure of the hypothetical *m*-$Sb_2Se_3$. The relative positions of atomic coordinate were kept unchanged and the cell volume was relaxed to reduce internal stress.

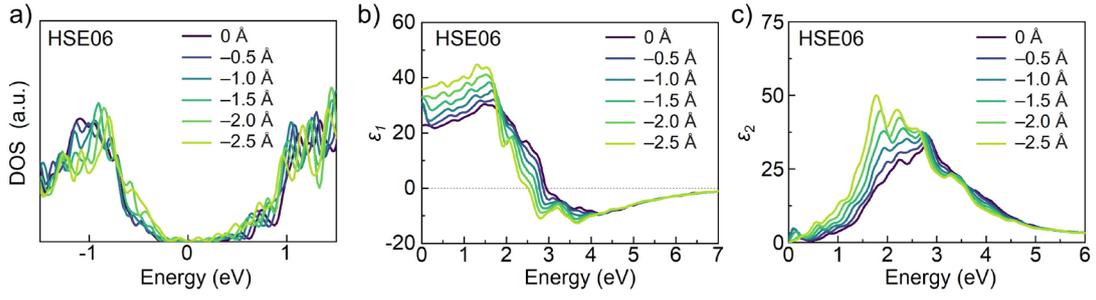

Fig. S3. The DFT-calculated a) DOS, b) real parts and c) imaginary parts of the dielectric functions for the strained $r$-Sb$_2$Se$_3$ models. The $c$ lattice length of the models is decreased gradually.

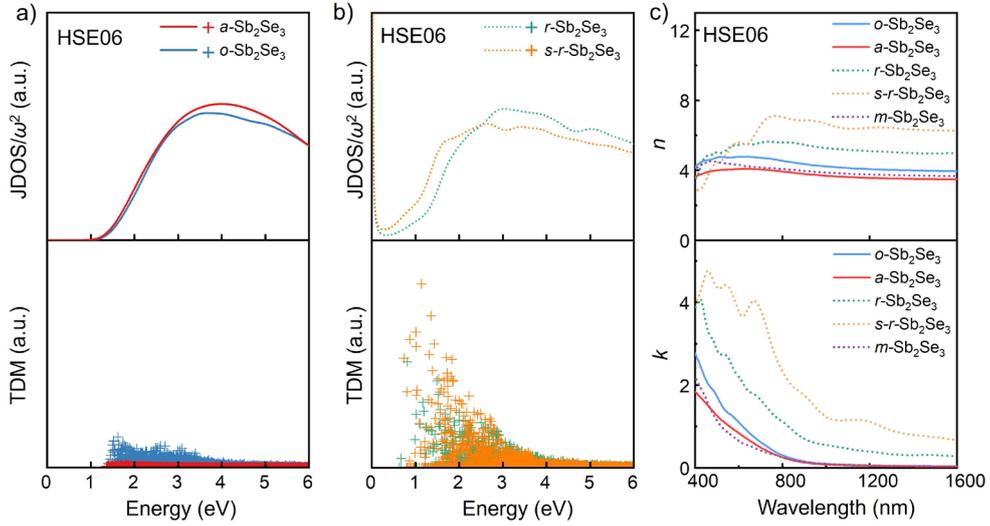

Fig. S4. The DFT-calculated a) Joint density of states (JDOS), b) transition dipole moment (TDM), c) $n$ and $k$ for $o$-, $r$-, $s$-$r$- and $a$-Sb$_2$Se$_3$.

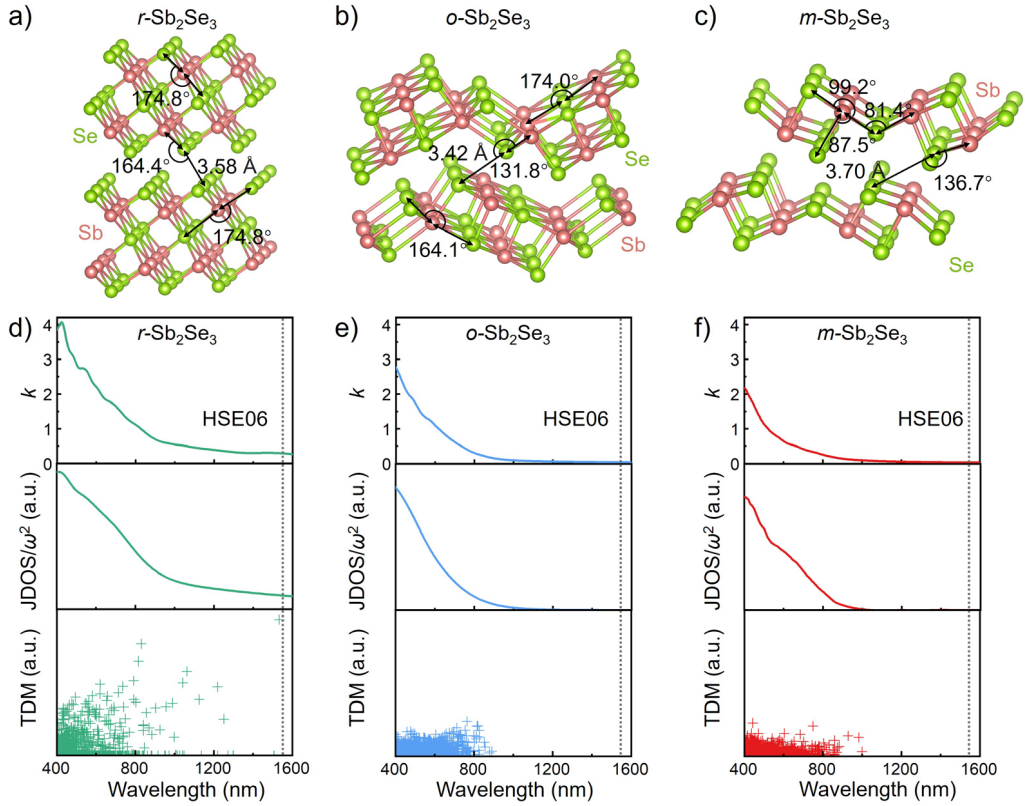

Fig. S5. DFT-calculated a) $r$-$Sb_2Se_3$, b) $o$-$Sb_2Se_3$, and c) $m$-$Sb_2Se_3$ crystalline models. The major bond angles are shown. The degree of $p$ orbital alignment gradually decreases. d-f) The extinction coefficient $k$, JDOS, and TDM calculated for d) $r$-$Sb_2Se_3$, e) $o$-$Sb_2Se_3$, and f) $m$-$Sb_2Se_3$. The dashed lines mark the wavelength of 1550 nm. For $r$-$Sb_2Se_3$, both the JDOS and TDM have non-negligible values, which together result in a finite optical absorption in the telecom band range. In contrast, the $o$- and $m$-$Sb_2Se_3$ models show neither JDOS nor TDM values, leading to the low-loss features in the telecom band range.

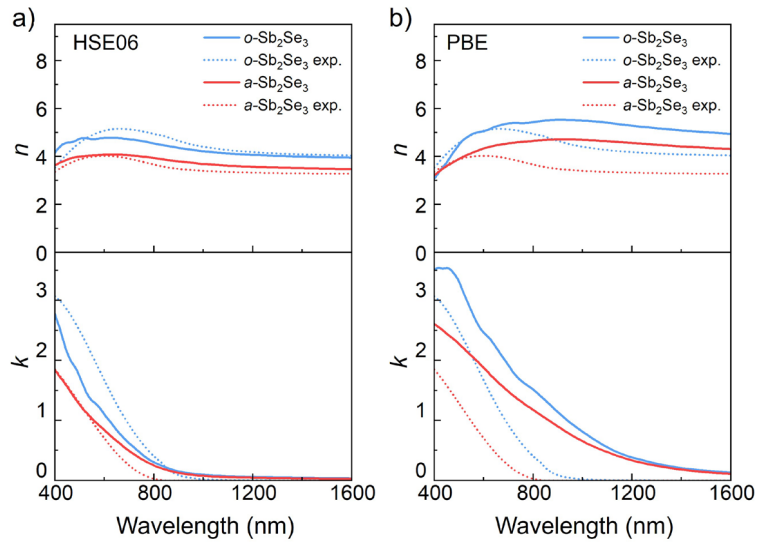

Fig. S6. The comparison between the experimental and calculated $n$/$k$ of amorphous and orthorhombic $Sb_2Se_3$. a) DFT calculations using the HSE06 hybrid functional. b) DFT calculations using the PBE functional. The experimental data are adapted from Ref.52.

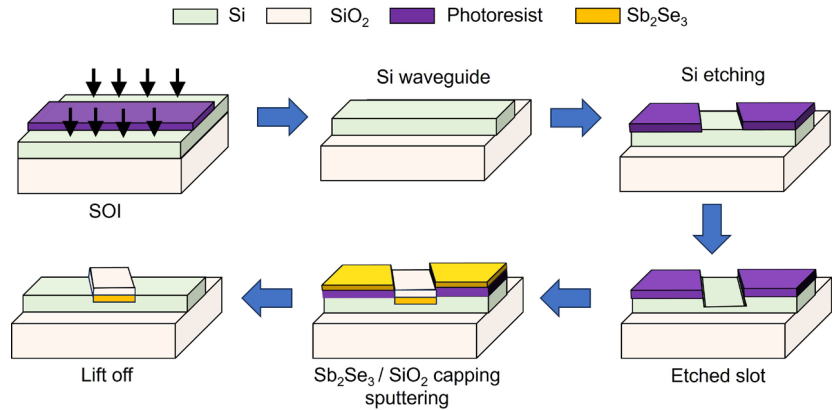

Fig. S7. A possible fabrication process of the $Sb_2Se_3$ PMC device. The primary challenge in device fabrication arises from the alignment deviation between $Sb_2Se_3$ and the etched air slot. By retaining the photoresist in silicon slot etching step as the sputtering window for subsequent deposition, it is able to avoid the misalignment caused by secondary lithography. Furthermore, by optimizing the lithography and etching processes, the variation can be controlled within ~10 nm, which has a negligible impact on device performance.

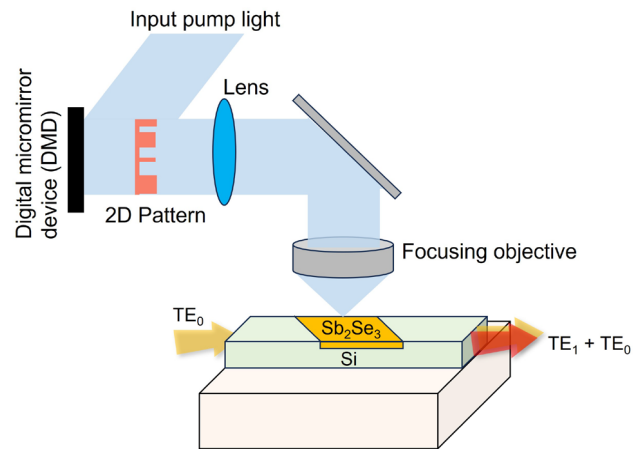

Fig. S8. Patterning of the $Sb_2Se_3$ PMC device using pump light and digital micromirror device[Ref.75]. By improving the accuracy of the linear stage, high-precision patterning could be possibly achieved.

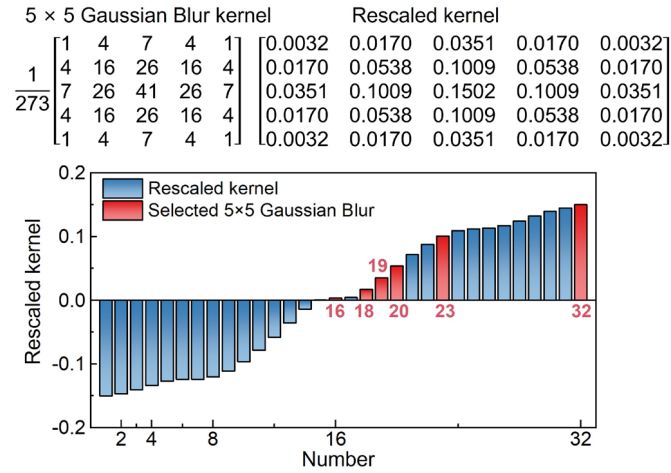

Fig. S9. The standard 5×5 Gaussian blur kernel and the rescaled kernel using PMC data. The selected mode contrast values were highlighted in red.

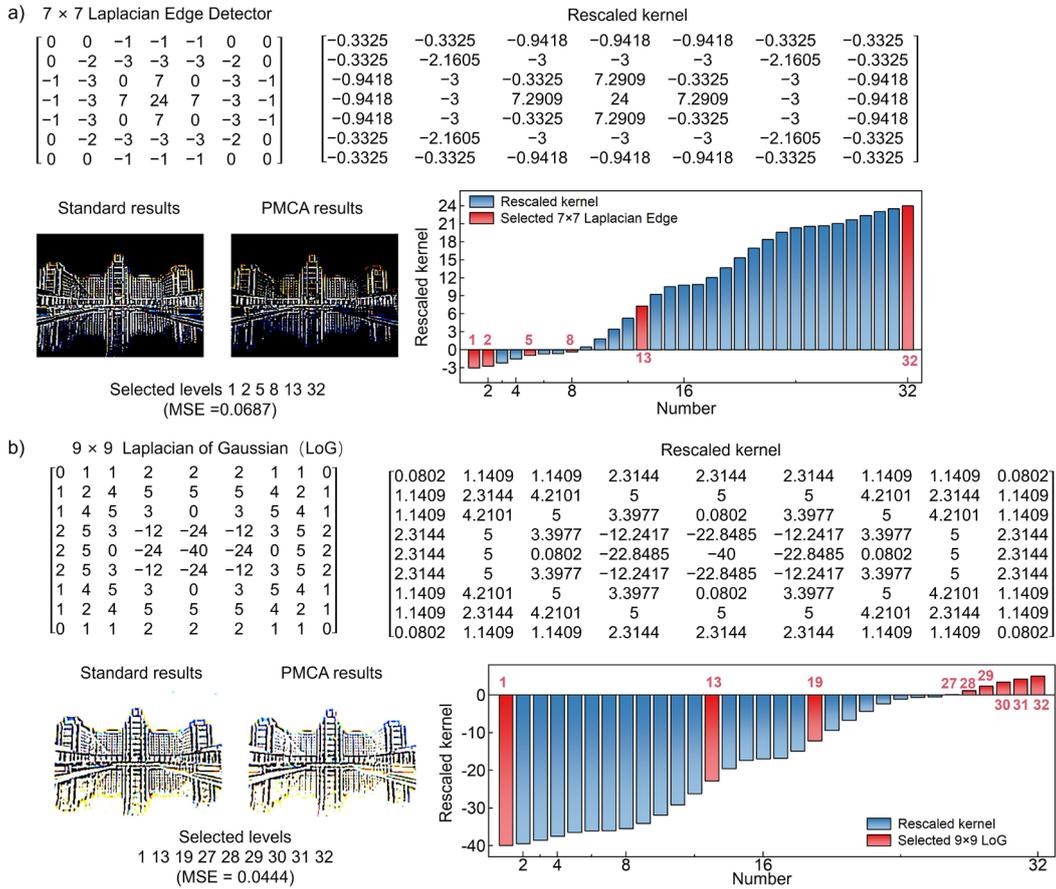

Fig. S10. The image processing results using the standard and PMC convolutional kernel values via a) 7×7 Laplacian edge detector kernel and b) 9×9 Laplacian of Gaussian (LoG) kernel.

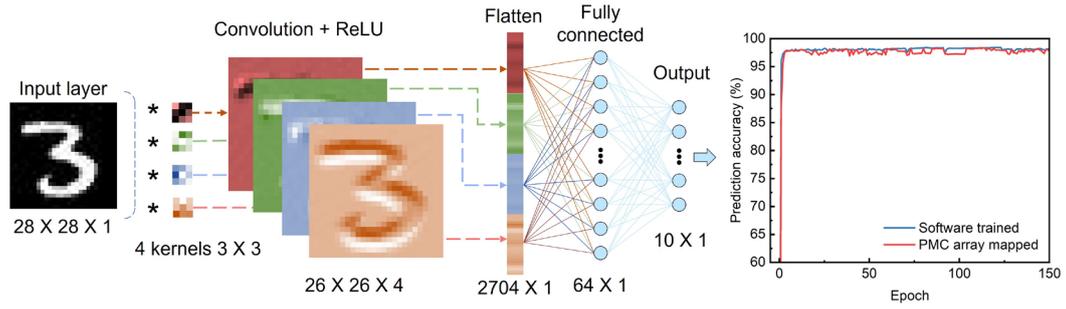

Fig. S11. The CNN framework based on the $Sb_2Se_3$ PMC array and the predicted recognition accuracy using the MINIST handwritten digits dataset.

| Dataset type | Year | Type of the platform | Prediction accuracy |
|---|---|---|---|
| MNIST fashion product | 2024[Ref. 38] | Partial coherence parallelized photonic tensor core convolutional network | 82.8% |
| | 2023[Ref. 32] | Photonic-electronic integrated convolutional network | 86.0% |
| | **This work** | **PMC convolutional network** | **87.6%** |
| MNIST handwritten digits | 2024[Ref. 38] | Partial coherence parallelized photonic tensor core convolutional network | 95.0% |
| | 2023[Ref. 32] | Photonic-electronic integrated convolutional network | 87.0% |
| | 2021[Ref. 33] | Integrated photonic tensor core convolutional network | 95.3% |
| | **This work** | **PMC convolutional network** | **97.8%** |

Table S1. Comparison of the MNIST fashion product/MNIST handwritten digits datasets and predicted recognition accuracies with various PCM-based photonic neural networks.